\documentclass{webofc}

\usepackage[varg]{txfonts}   
\usepackage{graphicx}
\usepackage{bm}
\usepackage{amssymb,amsmath,latexsym}
\usepackage{color}
\definecolor{darkblue}{RGB}{0,0,196}
\definecolor{darkred}{RGB}{196,0,0}
\usepackage[colorlinks=true,linktocpage=true,linkcolor=darkblue,citecolor=darkblue,urlcolor=darkblue]{hyperref}
\usepackage{array}

\begin{document}

\title{Non-equilibrium evolution of quarkonium in medium in the open quantum system approach}

\author{\firstname{Michael} \lastname{Strickland}\inst{1}\fnsep\thanks{\email{mstrick6@kent.edu}}}

\institute{Department of Physics, Kent State University, Kent, OH 44242, USA}

\abstract{
In this proceedings contribution, I review recent work that aims to provide a more comprehensive and systematic understanding of bottomonium dynamics in the quark-gluon plasma using an open quantum system (OQS) approach that is applied in the framework of the potential non-relativistic QCD (pNRQCD) effective field theory and coupled to realistic hydrodynamical backgrounds that have been tuned to soft hadron observables.  I review how the computation of bottomonium suppression can be reduced to solving a Gorini-Kossakowski-Sudarshan-Lindblad (GKSL) equation for the evolution of the $b\bar{b}$ reduced density matrix, which includes both singlet and octet states plus medium-induced transitions between them at next-to-leading order (NLO) in the binding energy over temperature.  Finally, I present comparisons of phenomenological predictions of the NLO OQS+pNRQCD approach and experimental data for bottomonium suppression and elliptic flow in LHC 5.02 TeV Pb-Pb collisions obtained using both smooth and fluctuating hydrodynamic initial conditions.
}

\maketitle

\section{Introduction}
\label{intro}

The experimentally-observed suppression of bottomonium production in heavy-ion collisions provides strong evidence for the creation of a deconfined quark-gluon plasma (QGP)~\cite{STAR:2013kwk,PHENIX:2014tbe,STAR:2016pof,CMS:2017ycw,Sirunyan:2018nsz,ALICE:2019pox,Acharya:2020kls,Lee:2021vlb,CMS:2020efs,CMS:2022rna}.
In the seminal works of Karsch, Mehr, Matsui, and Satz \cite{Matsui:1986dk,Karsch:1987pv}, heavy-quarkonium suppression was proposed as a signal of the formation of a deconfined QGP.  In these early works the suppression of heavy-quarkonium production was conjectured to be due to Debye screening of chromoelectric fields, which resulted in a modification of the heavy-quark heavy-anti-quark potential.  In recent years it was shown that, when computed systematically, in addition to Debye screening of the real part of the potential there also exists an imaginary contribution to the  potential, which results in there being large in-medium widths for heavy-quarkonium bound states~\cite{Laine:2006ns,Brambilla:2008cx,Beraudo:2007ky,Escobedo:2008sy,Dumitru:2009fy,Brambilla:2010vq,Brambilla:2011sg,Brambilla:2013dpa}.  
The imaginary part of the potential has its origin primarily in two effects: Landau damping, which is related to parton dissociation \cite{Brambilla:2013dpa}, and singlet to octet transitions, which is an effect particular to quantum chromodynamics (QCD) \cite{Brambilla:2008cx} and is related to the gluo-dissociation process \cite{Brambilla:2011sg}.  
At temperatures relevant in current heavy-ion collision experiments, the imaginary part of the potential results in widths that can be on the order of 100's of MeV, meaning that the decays occur on the timescales relevant for QGP dynamics and must therefore be taken into account in phenomenological applications.
 
\begin{figure}[t]
\begin{center}
\includegraphics[width=0.95\linewidth]{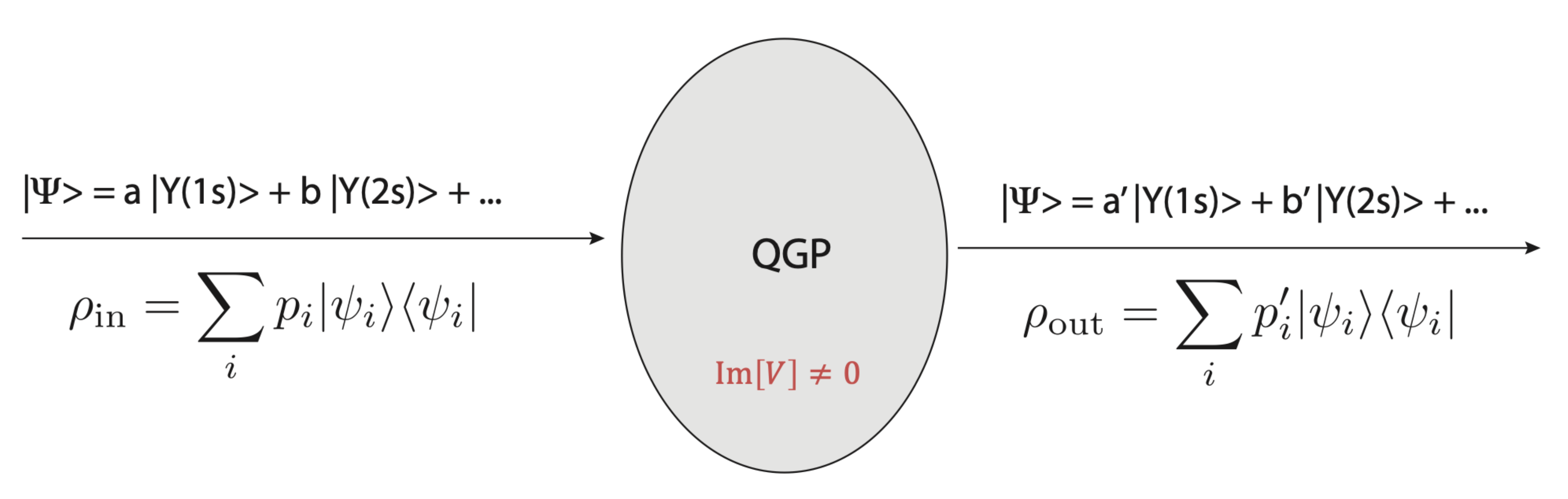} 
\end{center}
\vspace{-5mm}
\caption{
Schematic representation of the modification of the heavy quark-antiquark wave function resulting from interactions with the QGP.
}
\label{fig:sketch}
\end{figure}

In recent years resummed perturbative and effective field theory calculations of the imaginary part of the heavy-quark potential \cite{Laine:2006ns,Brambilla:2008cx,Beraudo:2007ky,Escobedo:2008sy,Dumitru:2009fy,Brambilla:2010vq,Brambilla:2011sg,Brambilla:2013dpa} have been confirmed by first-principles non-perturbative Euclidean lattice QCD and classical real-time lattice QCD measurements~\cite{Rothkopf:2011db,Petreczky:2018xuh,Rothkopf:2019ipj,Bala:2019cqu,Laine:2007qy,Lehmann:2020fjt,Boguslavski:2020bxt}.  Also, starting over a decade ago, phenomenological calculations of heavy-quarkonium suppression have included this effect~\cite{Strickland:2011mw,Strickland:2011aa,Krouppa:2015yoa,Islam:2020bnp,Islam:2020gdv,Brambilla:2020qwo,Brambilla:2021wkt,Brambilla:2022ynh,Wen:2022yjx,Alalawi:2022gul}.  These studies have provided strong indications that a self-consistent quantum mechanical description of heavy-quarkonium evolution in the QGP is now possible.  One of the conceptual issues brought forth by such studies was how to formulate the time evolution of the reduced density matrix of the heavy quark-antiquark pairs in a manner that preserved unitary and positivity.  The solution to this problem came through the formulation of the problem as an open quantum system (OQS) in which heavy quark-antiquark pairs are described in terms of a reduced density matrix obtained by integrating out the medium degrees of freedom \cite{Breuer:2002pc,Akamatsu:2020ypb,Yao:2021lus}.  In Fig.~\ref{fig:sketch}, I sketch the conceptual setup implied by such studies, wherein a quantum linear superposition expressed either in terms of the wave function or the reduced density matrix must be evolved through a time-evolving QGP.  From such a calculation one extracts the survival probabilities of quarkonium states from the probability to find a given eigenstate in the final quantum state relative to the probability to find it in the initial quantum state.

As a result of these calculations, in the last decade there has been a fundamental paradigm shift in our treatment of heavy-quarkonium dynamics and many important works on the application of OQS methods to theoretical models of heavy-quarkonium suppression have emerged~\cite{Akamatsu:2011se,Akamatsu:2014qsa,Blaizot:2015hya,Brambilla:2016wgg,Blaizot:2017ypk,Brambilla:2017zei,Blaizot:2018oev,Yao:2018nmy,Miura:2019ssi,Brambilla:2019tpt,Sharma:2019xum,Akamatsu:2020ypb,Yao:2020xzw,Yao:2020eqy,Blaizot:2021xqa,Yao:2021lus,Katz:2015qja}.
Here I will focus on recent works by myself and my collaborators that apply OQS methods within the framework of the potential non-relativistic QCD (pNRQCD) effective field theory (EFT) \cite{Pineda:1997bj,Brambilla:1999xf,Brambilla:2004jw}.
The pNRQCD EFT relies on there being a large separation between the energy scales in the problem, which is guaranteed for systems where the velocity of the heavy quark relative to the center of mass is small, i.e., $v \ll 1$, and such EFTs have been extended to study quarkonium at finite temperature~\cite{Brambilla:2008cx,Escobedo:2008sy,Brambilla:2010vq,Brambilla:2011sg,Brambilla:2013dpa}.  In Refs.~\cite{Brambilla:2016wgg,Brambilla:2017zei,Brambilla:2019tpt}, the authors considered, among other possibilities, the scale hierarchy relevant for small bound states in a high-temperature QGP, $1/r \sim Mv \gg m_D \sim \pi T \gg E$, where $r$ is the typical size of the state, $M$ is the heavy quark mass, $m_D$ is the Debye mass, $T$ is the temperature, and $E$ is the binding energy.  

Based on the scale hierarchy above, the authors of Ref.~\cite{Brambilla:2019tpt} derived a Gorini-Kossakowski-Sudarshan-Lindblad (GKSL) equation~\cite{Gorini:1975nb,Lindblad:1975ef} for the heavy-quarkonium reduced density matrix. Recently, the \texttt{QTraj} code of Ref.~\cite{Omar:2021kra}, which implements a Monte Carlo quantum trajectories algorithm~\cite{Dalibard:1992zz} for solving GKSL-type equations, was used to make phenomenological predictions for various heavy-ion collision bottomonium observables~\cite{Brambilla:2020qwo,Brambilla:2021wkt,Brambilla:2022ynh}.  
For this purpose, the GKSL solver was coupled to a 3+1D viscous hydrodynamics code, which used smooth (optical) Glauber initial conditions \cite{Alqahtani:2017mhy,Alqahtani:2020paa,Alalawi:2021jwn}.  Most recently, the OQS+pNRQCD framework was extended to next-to-leading order (NLO) in the binding energy over the temperature, allowing it to be applied at lower temperatures than the leading-order formalism~\cite{Brambilla:2022ynh}.  In addition, the effect of fluctuating initial conditions in the hydrodynamic background was investigated using the NLO OQS+pNRQCD framework in Ref.~\cite{Alalawi:2022gul}.
In Refs.~\cite{Brambilla:2022ynh,Alalawi:2022gul} it was found that the NLO OQS+pNRQCD framework provided a good description of existing experimental data from the ALICE, ATLAS, and CMS collaborations for both the nuclear modification factor, $R_{AA}$, and elliptic flow, $v_2$.

\section{Methodology}

In Refs.~\cite{Brambilla:2016wgg,Brambilla:2017zei} the evolution of the heavy-quarkonium reduced density matrix in the regime $Mv \gg T$ was obtained.
The resulting evolution equations were
\begin{align}
	\frac{d\rho_{s}(t)}{dt} &= -i[h_{s}, \rho_{s}(t)] - \Sigma_{s} \rho_{s}(t) - \rho_{s}(t) \Sigma^{\dagger}_{s} + \Xi_{so}(\rho_{o}(t)) \, ,\\
	\frac{d\rho_{o}(t)}{dt} &= -i[h_{o}, \rho_{o}(t)] - \Sigma_{o} \rho_{o}(t) - \rho_{o}(t) \Sigma_{o}^{\dagger}
	+ \Xi_{os}(\rho_{s}(t)) + \Xi_{oo}(\rho_{o}(t)) \, ,
\end{align}
where $\rho_s$ and $\rho_o$ are the singlet and octet heavy-quarkonium reduced density matrices and no relation between the temperature $T$ and the binding energy $E$ has yet been assumed.  In the large time limit, one has
\begin{align}
	\Sigma_{s} &= r_{i} A_{i}^{so \dagger}, \\
	\Sigma_{o} &= \frac{1}{N_{c}^{2}-1} r_{i} A_{i}^{os \dagger}
	+ \frac{N_{c}^{2}-4}{2(N_{c}^{2}-1)}r_{i}A_{i}^{oo \dagger}, \\
	\Xi_{so}(\rho_{o}(t)) &= \frac{1}{N_{c}^{2}-1} \left( A_{i}^{os \dagger} \rho_{o}(t) r_{i} + r_{i} \rho_{o}(t) A_{i}^{os}\right), \\
	\Xi_{os}(\rho_{s}(t)) &= A_{i}^{so \dagger} \rho_{s}(t) r_{i} 
	+ r_{i} \rho_{s}(t) A_{i}^{so}, \\
	\Xi_{oo}(\rho_{o}(t)) &= \frac{N_{c}^{2}-4}{2(N_{c}^{2}-1)}\left(
	A_{i}^{oo \dagger} \rho_{o}(t) r_{i} + r_{i} \rho_{o}(t) A_{i}^{oo}\right),
\end{align}
with
\begin{align}
	A_{i}^{uv} &= \frac{g^{2}}{6N_{c}} \int^{\infty}_{0} \text{d}s\, e^{-i h_{u}s} r_{i} e^{i h_{v} s} 
	\langle \tilde{E}^{a}_j(0, {\bm 0}) \tilde{E}^{a}_j(s, {\bm 0})\rangle \, ,\label{eq:auv_operator}\\
	\tilde{E}^{a}_i(s, {\bm 0}) &= \Omega(s)^\dagger E^{a}_i(s, {\bm 0}) \Omega(s) \, ,\\
	\Omega(s) &= \text{exp}\left[  -ig \int_{-\infty}^{s} \text{d}s' A_{0}(s', {\bm 0}) \right].
\end{align}
Above, $i \in \{x,y,z\}$, $g$ is the strong coupling, $A_0$ is the temporal component of the gauge field, $E_i^a$ are chromoelectric fields, $N_{c}$ is the number of colors, and $h_{u,v}$ represents either the singlet or octet Hamiltonian,
$h_{s,o}={\bm p}^{2}/M+V_{s,o}$, with \mbox{$V_{s}= -C_F\alpha_{\rm s}/r$} and $V_{o}=\alpha_{\rm s}/2N_{c}r$.  In the potentials, $C_F=(N_{c}^{2}-1)/2N_{c}$ is the Casimir of the fundamental representation and the strong coupling constant $\alpha_{\rm s} = g^2/4\pi$ should be evaluated at the inverse Bohr radius $1/a_{0}$.

\subsection{The quantum master equation for heavy-quarkonium evolution}

These equations can be written as a  quantum master equation
\begin{equation}\label{eq:master_equation}
	\frac{d\rho(t)}{dt} = -i \left[ H, \rho(t) \right] + \sum_{nm} h_{nm} \left( L_{i}^{n} \rho(t) L^{m\dagger}_{i} - \frac{1}{2} \left\{ L^{m\dagger}_{i} L_{i}^{n}, \rho(t) \right\} \right),
\end{equation}
where
\begin{equation}\label{eq:master_equation_rho_and_h}
	\rho(t) = \begin{pmatrix} \rho_{s}(t) & 0 \\ 0 & \rho_{o}(t) \end{pmatrix} \,,\text{\quad} H = \begin{pmatrix} h_{s} + \text{Im}(\Sigma_{s}) & 0 \\ 0 & h_{o} + \text{Im}(\Sigma_{o}) \end{pmatrix} ,
\end{equation}
\begin{equation}\label{eq:master_equation_l0_l1}
	L_{i}^{0} = \begin{pmatrix} 0 & 0 \\ 0 & 1 \end{pmatrix}r_{i} \text{\,, \quad} L_{i}^{1} = \begin{pmatrix} 0 & 0 \\ 0 & \frac{N_{c}^{2}-4}{2(N_{c}^{2}-1)} A_{i}^{oo \dagger} \end{pmatrix},
\end{equation}
\begin{equation}\label{eq:master_equation_l2_l3}
	L_{i}^{2} = \begin{pmatrix} 0 & 1 \\ 1 & 0 \end{pmatrix}r_{i} \text{\,, \quad} L_{i}^{3} = \begin{pmatrix} 0 & \frac{1}{N_{c}^{2}-1} A_{i}^{os \dagger} \\ A_{i}^{so \dagger} & 0 \end{pmatrix},
\end{equation}
and
\begin{equation}\label{eq:metric_tensor}
	h = \begin{pmatrix} 0 & 1 & 0 & 0 \\ 1 & 0 & 0 & 0 \\ 0 & 0 & 0 & 1 \\ 0 & 0 & 1 & 0 \end{pmatrix}.
\end{equation}
Due to the fact that, for general $T$ and $E$,  $h$ possesses negative eigenvalues, Eq.~(\ref{eq:master_equation}) can result in negative probabilities.  This is similar to the widely used Caldeira--Leggett quantum master equation \cite{Caldeira:1982iu}.  However, when expanding in powers of the binding energy over the temperature, it is possible to obtain an evolution equation that maintains positivity by discarding terms of higher order in $E/T$.  Using such a strategy, in Ref.~\cite{Brambilla:2022ynh} we derived a GKSL equation that is accurate to order $E/T$ and respects positivity.\footnote{A similar strategy was employed in section of 4 of Ref.~\cite{Akamatsu:2020ypb}.}  This allows one to map the quantum master equation to a GKSL equation that can be solved using a stochastic unraveling called the quantum trajectories algorithm~\cite{Omar:2021kra,Brambilla:2022ynh}.

\subsection{Reduction to a GKSL equation at NLO in the binding energy over temperature}

In the high temperature limit, one can expand the exponentials appearing in the quantum master equation, retaining terms up to order $E/T$.  The resulting $A_{i}^{uv}$ takes the form
\begin{align}\label{eq:nlo_medium_interaction}
	A_{i}^{uv} =& \frac{r_{i}}{2} (\kappa - i \gamma) + \kappa \left( -\frac{i p_{i}}{2MT} + \frac{\Delta V_{uv}}{4T} r_{i} \right) + \cdots,
\end{align}
where $p_i$ is the momentum operator, $\Delta V_{uv}$ is the difference of the singlet or octet potentials, and we have discarded terms of order $(E/T)^{2}$ or higher.
Above, the coefficient $\kappa$ is the heavy-quarkonium momentum-diffusion coefficient and $\gamma$ is its dispersive counterpart
\begin{align}
	\kappa &= \frac{g^{2}}{6N_{c}} \int_{0}^{\infty}\text{d}s\, \left\langle \left\{\tilde{E}^{a}_i(s,\vec{0}), \tilde{E}^{a}_i(0,\vec{0})\right\}\right\rangle ,\label{eq:kappa}\\ 
	\gamma &= -i\frac{g^{2}}{6N_{c}} \int_{0}^{\infty}\text{d}s\, \left\langle \left[\tilde{E}^{a}_i(s,\vec{0}), \tilde{E}^{a}_i (0,\vec{0})\right]\right\rangle.\label{eq:gamma}
\end{align}

Using this expression, the operators $L_i^1$ and $L_i^3$ in the quantum master equation can be expressed as
\begin{align}
	&L_{i}^{1} = \frac{N_{c}^{2}-4}{2(N_{c}^{2}-1)} \begin{pmatrix} 0 & 0 \\ 0 &  1 \end{pmatrix} \left[ \frac{r_{i}}{2} (\kappa + i \gamma) + \kappa \frac{i p_{i}}{2MT} \right] ,
	\label{eq:Li1NLO}\\
	&\begin{aligned} L_{i}^{3}=
	&\begin{pmatrix} 0 & \frac{1}{N_{c}^{2}-1} \\ 0 & 0 \end{pmatrix} \left[\frac{r_{i}}{2} (\kappa + i \gamma) + \kappa \left( \frac{i p_{i}}{2MT} + \frac{\Delta V_{os}}{4T} r_{i} \right) \right] \\
	& \hspace{1cm} + \begin{pmatrix} 0 & 0 \\ 1 & 0 \end{pmatrix} \left[\frac{r_{i}}{2} (\kappa + i \gamma) + \kappa \left( \frac{i p_{i}}{2MT} + \frac{\Delta V_{so}}{4T} r_{i} \right) \right] .
	\end{aligned}
\end{align}
The operators $L_i^0$ and $L_i^2$ are the same as in Eqs.~\eqref{eq:master_equation_l0_l1} and~\eqref{eq:master_equation_l2_l3}, respectively.  The resulting quantum master equation cannot be written in GKSL form since the $L^{n}_i$ operators are linearly independent \cite{Brambilla:2022ynh}; however, the operator vector $(L^{0}, L^{1}, L^{2}, L^{3})$ can be transformed such that only components of order $E/T$ project on the eigenspaces of the negative eigenvalues of the matrix $h_{nm}$.  Neglecting terms of order $(E/T)^{2}$, one can justifiably discard such contributions.

The resulting GKSL equation, accurate to order $E/T$, is
\begin{equation}
\frac{d\rho(t)}{dt} = -i \left[ H, \rho(t) \right] + \sum_{n=0}^1
\left( C_{i}^{n} \rho(t) C^{n\dagger}_{i} - \frac{1}{2} \left\{ C^{n \dagger}_{i} C_{i}^{n}, \rho(t) \right\} \right),
\label{eq:Lindblad}
\end{equation}
with the Hamiltonian being
\begin{equation}
	H = \begin{pmatrix} h_{s} + \text{Im}(\Sigma_{s}) & 0 \\ 0 & h_{o} + \text{Im}(\Sigma_{o}) \end{pmatrix} .
\end{equation}
At order $E/T$, the self-energies appearing above can be expressed as
\begin{align}\label{eq:self_energies_im}
	\text{Im}\left( \Sigma_{s} \right) &= \frac{r^{2}}{2} \gamma +\frac{\kappa}{4MT} \{r_{i}, p_{i}\} \, , \\
	\text{Im}\left( \Sigma_{o} \right) &= \frac{N_{c}^{2}-2}{2(N_{c}^{2}-1)} \left( \frac{r^{2}}{2} \gamma +\frac{\kappa}{4MT} \{r_{i}, p_{i}\} \right) ,
\end{align}
and the collapse operators as
\begin{align}
	    C_{i}^{0} =& \sqrt{\frac{\kappa}{N_{c}^{2}-1}} \begin{pmatrix} 0 & 1 \\ 0 & 0 \end{pmatrix} \left(r_{i} + \frac{i p_{i}}{2MT} +\frac{\Delta V_{os}}{4T}r_{i} \right) + \sqrt{\kappa} \begin{pmatrix} 0 & 0 \\ 1 & 0 \end{pmatrix} \left(r_{i} + \frac{i p_{i}}{2MT} +\frac{\Delta V_{so}}{4T}r_{i} \right), \\
	C_{i}^{1} &= \sqrt{\frac{\kappa(N_{c}^{2}-4)}{2(N_{c}^{2}-1)}} \begin{pmatrix} 0 & 0 \\ 0 & 1 \end{pmatrix} \left(r_{i} + \frac{i p_{i}}{2MT} \right).\label{eq:c1}
\end{align}

\subsection{Hamiltonian and width operators in the reduced wave function basis}

The decay width of a state can be obtained by taking the trace of the anti-commutator term in the GKSL equation~\eqref{eq:Lindblad}
\begin{equation}
    \Gamma \left[
    \rho_{u}^{l}(r)
    \right]\equiv \sum_{n=0}^1
    \text{Tr} \left[ C^{n\,\dagger} C^{n} \rho^{l}_{u}(r) \right],
\end{equation}
where $\rho^{l}_{u}(r)$ is the state being considered and $\Gamma[ \rho_{u}^{l}(r) ]$ is a functional returning the width.  The trace on the right-hand-side is over color, angular momentum, and position.  When expressed as operators acting on the reduced wave function, $u(t,r) \equiv r R(t,r)$, the singlet-octet and octet-octet width operators take the form~\cite{Brambilla:2022ynh}
\begin{align}
    \overline{\Gamma}_{o\to s}^{\uparrow} &= \frac{\hat\kappa T^3}{N_c^2-1} \frac{l+1}{2l+1} \left[ \left( r +  \frac{N_c \alpha_{\rm s}}{8T} \right)^2 - \frac{3}{2MT} + \frac{\overline{{\cal D}}^2}{(2 M T)^2} - \frac{1}{2 M T} \left( \frac{N_c \alpha_{\rm s}}{4T} \right) \frac{1}{r} \right] , \label{eq:gupos} \\
    \overline{\Gamma}_{o\to s}^{\downarrow} &= \frac{l}{l+1} \overline{\Gamma}_{o\to s}^{\uparrow} \, , \\
    \overline{\Gamma}_{o\to o}^{\uparrow} &=  \hat\kappa T^3 \frac{N_c^2-4}{2(N_c^2-1)} \frac{l+1}{2l+1} \left[ r^2 - \frac{3}{2MT} + \frac{\overline{{\cal D}}^2}{(2 M T)^2} \right] \, , \\
    \overline{\Gamma}_{o\to o}^{\downarrow} &= \frac{l}{l+1}  \overline{\Gamma}_{o\to o}^{\uparrow} \, , \label{eq:gdownoo}
\end{align}
with $\overline{{\cal D}}^2 \equiv -\partial^2/\partial r^2 + l (l+1)/r^2$.  The singlet-octet width operator is not required since singlet states must always transition to an octet state and the outgoing angular momentum quantum number can be chosen using $\overline{\Gamma}^\downarrow_{s \to o} / \overline{\Gamma}^\uparrow_{s \to o} = l/(l+1)$~\cite{Brambilla:2022ynh}.

The effective Hamiltonians for singlet and octet evolution can be obtained using the relations $H^{\rm eff}_{s,o} = h_{s,o}  + \text{Im}(\Sigma_{s,o})  - i\Gamma_{s,o}/2$ with $\Gamma_s = \sum_{i \in \{\uparrow,\downarrow\}} \Gamma_{s\rightarrow o}^i$ and $\Gamma_o =  \sum_{i \in \{\uparrow,\downarrow\}} (\Gamma_{o\rightarrow s}^i + \Gamma_{o\rightarrow o}^i$).  
Expressing the Hamiltonians as operators acting on the reduced wave function $u$, the singlet effective Hamiltonian $\overline{H}^{\rm eff}_s$ is given by~\cite{Brambilla:2022ynh}
\begin{align}
    \text{Re}[\overline{H}^{\rm eff}_s] &=  
    \frac{\overline{\cal D}^2}{M}
    - \frac{C_F\, \alpha_{\rm s}}{r} + \frac{\hat\gamma T^3}{2} r^2 +  \frac{\hat\kappa T^2}{4 M} \{r,p_r\} \, , \label{eq:heff1}\\
    \text{Im}[\overline{H}^{\rm eff}_s] &= - \frac{\hat\kappa T^3}{2} \left[ \left( r - \frac{N_c \alpha_{\rm s}}{8T} \right)^2 - \frac{3}{2MT} + \frac{{\overline{\cal D}}^2}{(2MT)^2} + \frac{1}{2MT} \left( \frac{N_c \alpha_{\rm s}}{4 T} \right)  \frac{1}{r}\right] ,
\end{align}
where $\hat{\kappa} = \kappa / T^3$, $\hat{\gamma} = \gamma / T^3$, and $p_r = - i \partial_r$.  The octet effective Hamiltonian $\overline{H}^{\rm eff}_o$ in the reduced wave function basis is given by
\begin{align}
    \text{Re}[\overline{H}^{\rm eff}_o] &=  \frac{\overline{\mathcal{D}}^{2}}{M} + \frac{1}{2N_{c}} \frac{\alpha_{\rm s}}{r} + \frac{N_c^2 - 2}{2(N_c^2 - 1)}\left[ \frac{ \hat\gamma T^3 }{2} r^2 +  \frac{\hat\kappa T^2}{4 M} \{r,p_r\} \right]  , \\
    \text{Im}[\overline{H}^{\rm eff}_o] &= - \frac{\hat\kappa T^3}{2(N_c^2-1)} \left[ \left( r + \frac{N_c \alpha_{\rm s}}{8T} \right)^2 - \frac{3}{2MT} + \frac{{\overline{\cal D}}^2}{(2MT)^2} - \frac{1}{2MT} \left( \frac{N_c \alpha_{\rm s}}{4 T} \right)  \frac{1}{r}\right]  \nonumber \\
&~ \hspace{1cm} {-}\frac{\hat\kappa T^3}{4(N_c^2-1)} \left[ r^2  
- \frac{3}{2MT} + \frac{{\overline{\cal D}}^2}{(2MT)^2} 
 \right] .  \label{eq:heff4}
\end{align}

\section{Results}

In Ref.~\cite{Brambilla:2022ynh} we considered $5.02$ TeV Pb-Pb collisions with the background temperature evolution given by 3+1D quasiparticle anisotropic hydrodynamics~\cite{Alqahtani:2020paa,Alalawi:2021jwn}.  For solving the GKSL equation, we ignored dynamical quantum jumps and evolved with the singlet effective Hamiltonian.  This was shown in prior works by myself and collaborators to be a quite reliable approximation in QCD due to the fact that the octet potential is repulsive \cite{Brambilla:2020qwo,Brambilla:2021wkt,Escobedo:2020tuc}.\footnote{Refs.~\cite{Brambilla:2020qwo,Brambilla:2021wkt} presented extensive calculations including the effect of dynamical jumps.} For the real-time quantum evolution, we used the Crank--Nicolson method on a one-dimensional lattice with $\texttt{NUM}=2048$ points and $\texttt{L}=40\,\mathrm{GeV}^{-1}$. The evolution time step was taken to be $\texttt{dt}=0.001\,\mathrm{GeV}^{-1}$.

The physical parameters necessary to proceed are the heavy quark mass $M$, the strong coupling $\alpha_{\rm s}$, and the transport coefficients $\kappa$ and $\gamma$.  For bottom quark mass we took $M = m_{b} = m_{\Upsilon(1S)}/2 = 4.73$ GeV with $m_{\Upsilon(1S)}$ from the Particle Data Group~\cite{Zyla:2020zbs}.  The Bohr radius $a_0$ was obtained from the 1-loop relation $a_{0} = 2/(C_F \, \alpha_{\rm s}(1/a_0)\,m_{b})$.  We evaluated $\alpha_{\rm s}$ at the inverse of the Bohr radius using the 1-loop running with $N_{f}=3$ flavors and $\Lambda_{\overline{\rm MS}}^{N_f=3}=332$ MeV~\cite{Zyla:2020zbs}, which resulted in $\alpha_{\rm s}(1/a_0) = 0.468$.  The transport coefficients $\kappa$ and $\gamma$ were fixed from direct and indirect lattice measurements, respectively.
The heavy-quark momentum-diffusion coefficient \cite{CasalderreySolana:2006rq,CaronHuot:2007gq} was measured directly in a quenched lattice simulation over a large range of temperatures in Ref.~\cite{Brambilla:2020siz}. 
We performed \texttt{QTraj} simulations using three parameterizations of $\hat{\kappa}(T)$, which we denote $\hat{\kappa}_{L}(T)$, $\hat{\kappa}_{C}(T)$, and $\hat{\kappa}_{U}(T)$.  These correspond to the lower, central, and upper ``fit'' curves shown in Fig.~13 of Ref.~\cite{Brambilla:2020siz}, respectively.
Since no direct lattice measurements of $\gamma$ exist at this moment in time, we made use of indirect estimates from unquenched lattice simulations.  In Ref.~\cite{Brambilla:2019tpt}, unquenched lattice measurements of the 1S mass shift, $\delta M[\Upsilon(1S)]$, from Refs.~\cite{Kim:2018yhk,Aarts:2011sm} were used to constrain $\hat{\gamma}(T) = \gamma / T^{3}$ between approximately  $-3.5 \leq \hat\gamma \leq 0$.  More recent unquenched lattice studies \cite{Larsen:2019bwy,Shi:2021qri} favor $\delta M(\Upsilon(1S)) \simeq 0$ and thus $\hat{\gamma} \simeq 0$.
We performed simulations using $\hat{\gamma}$ in the full range $-3.5 \leq \hat\gamma \leq 0$ and used this to estimate our systematic uncertainty due to $\gamma$.

To initialize the quantum evolution, we assumed that at $\tau = 0$ fm the bottomonium wave function was in a singlet state with a smeared delta function initial condition.  For this purpose, we assumed that the initial reduced wave function was given by a Gaussian delta function multiplied by a power of $r$ appropriate for the boundary conditions necessary for a given initial angular momentum $l$, i.e., $u_{\ell}(t_0) \propto r^{l+1} e^{-r^{2}/(ca_{0})^2}$, with $u$ normalized to one when summed over the entire (one-dimensional) lattice volume.  We took $c=0.2$, which is the same value as was used in our earlier works~\cite{Omar:2021kra}.  We evolved this initial wave function using the vacuum potential ($\hat\kappa=\hat\gamma=0$) from $\tau = 0$~fm to $\tau_{\rm med}$ = 0.6~fm and began in-medium evolution at $\tau =  \tau_{\rm med}$.  We terminated the quantum evolution when the local temperature along a given physical trajectory dropped below $T_F = 190$~MeV.  In each centrality bin, we sampled the initial production points from the binary collision overlap profile.  The initial transverse momenta were sampled from a $1/E_T^4$ spectrum and the azimuthal angles were sampled uniformly in $[0,2\pi)$.  We assumed that the wave packets traveled along eikonal trajectories and sampled the temperature along each trajectory generated.

From this evolution we computed the survival probability for each of the vacuum eigenstates by projecting the final time-evolved wave function with vacuum bottomonium eigenstates.  However, it was then necessary to take into account late time feed down of the excited states.  For this, we followed Ref.~\cite{Brambilla:2020qwo}, where a feed down matrix $F$ that relates the experimentally observed and direct $pp$ production cross sections cross sections, $\vec{\sigma}_{\text{exp}} = F \vec{\sigma}_{\text{direct}}$, was introduced.  The cross section vectors in this relation correspond to the cross-sections of the states considered and the matrix $F$ was fixed using the known branching fractions of excited bottomonium states \cite{Zyla:2020zbs}.  The states considered were $\vec{\sigma} = \{ \Upsilon(1S),\,$ $\Upsilon(2S),\,$ $\chi_{b0}(1P),\,$ $\chi_{b1}(1P),\,$ $\chi_{b2}(1P),\,$ $\Upsilon(3S),\,$ $\chi_{b0}(2P),\,$ $\chi_{b1}(2P),\,$ $\chi_{b2}(2P)\}$.  In practice, one has $F_{ij} = B_{j \rightarrow i}$ for $i<j$, $F_{ij}$ = 1 for $i=j$, and $F_{ij}$ = 0 for $i>j$, where $B_{j \rightarrow i}$ is the branching fraction of state $j$ into state $i$.  The specific values used can be found in Eq.~(6.4) of Ref.~\cite{Brambilla:2020qwo}.

Based on this, the nuclear modification factor $R_{AA}^i$ for bottomonium state $i$ is computed as
\begin{equation}
R^{i}_{AA}(c,p_T,\phi) = \Bigg\langle \frac{\left(F \cdot S(c,p_T,\phi) \cdot \vec{\sigma}_{\text{direct}}\right)^{i}}{\vec{\sigma}_{\text{exp}}^{i}} \Bigg\rangle \, ,
\label{eq:feeddown}
\end{equation}
where here $i$ labels the state being considered, $S(c,p_T,\phi)$ is the GKSL survival probability, $c$ labels the event centrality class, $p_T$ is the transverse momentum of the state, and $\phi$ its azimuthal angle. The direct production cross sections appearing in Eq.~\eqref{eq:feeddown} were obtained by inverting the feeddown matrix, $\vec{\sigma}_{\text{direct}} = F^{-1} \vec{\sigma}_{\text{exp}}$.   The angle brackets indicate a double average over (a) all physical trajectories of bottomonium states in the centrality and $p_T$ bin considered and (b) the hydrodynamic initial conditions used in each centrality bin.  Note that, for smooth initial conditions, the second average can be ignored.  For the integrated experimental cross sections we refer the reader to Sec.~6.4 of Ref.~\cite{Brambilla:2020qwo}.  To obtain $v_2$ in each centrality class, we computed $\langle \cos(2(\phi-\Psi_2) \rangle_{i,c,p_T}$, where the average is over all bottomonium states of type $i$ produced in the corresponding centrality and transverse momentum bins and $\Psi_2$ is second-order event plane angle, which is determined self-consistently from the final-state charged hadrons on an event-by-event basis.  For smooth initial conditions, $\Psi_2 = 0$ for all events.

\begin{figure}[t]
\begin{center}
\includegraphics[width=0.48\linewidth]{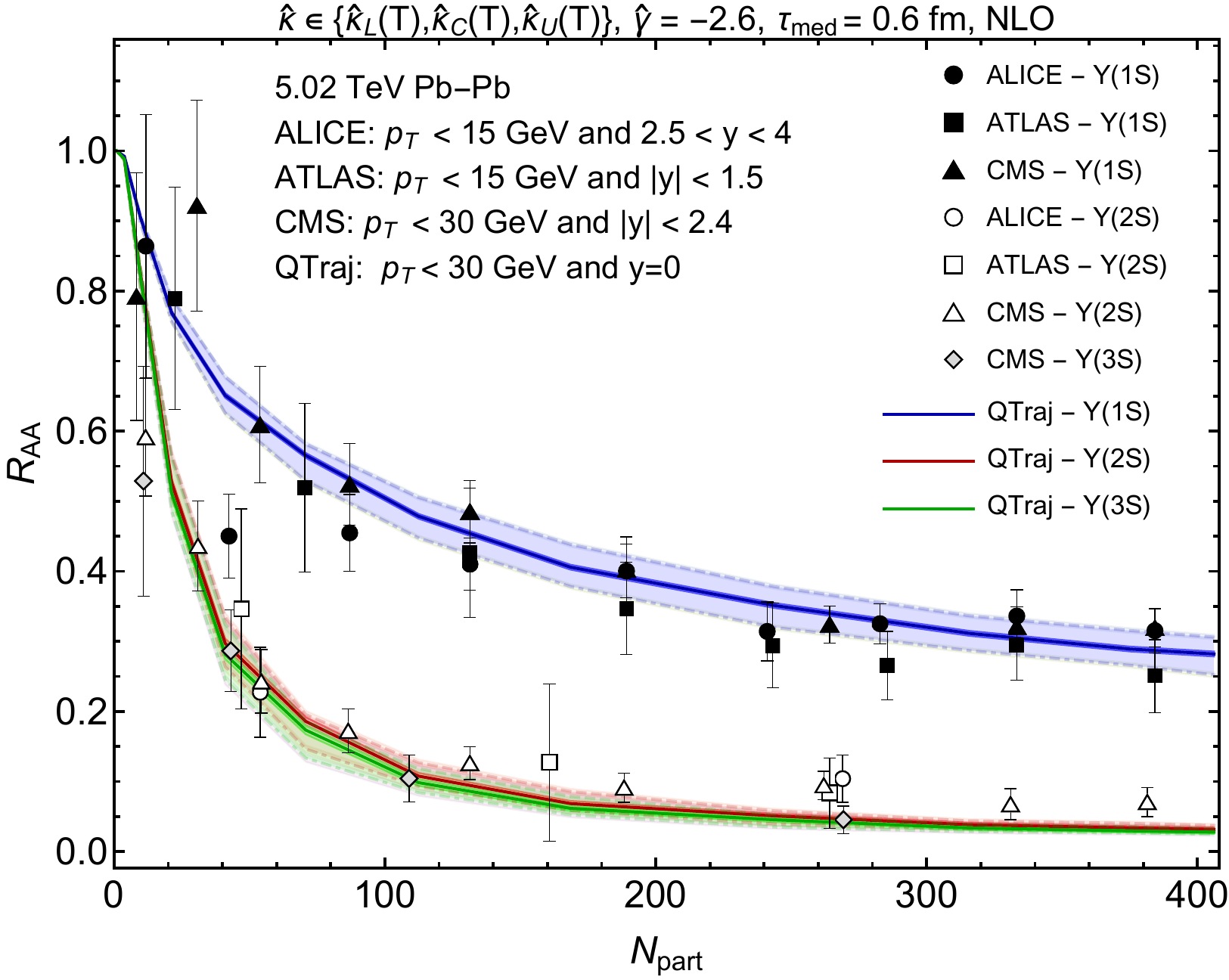} \hspace{2mm}
\includegraphics[width=0.48\linewidth]{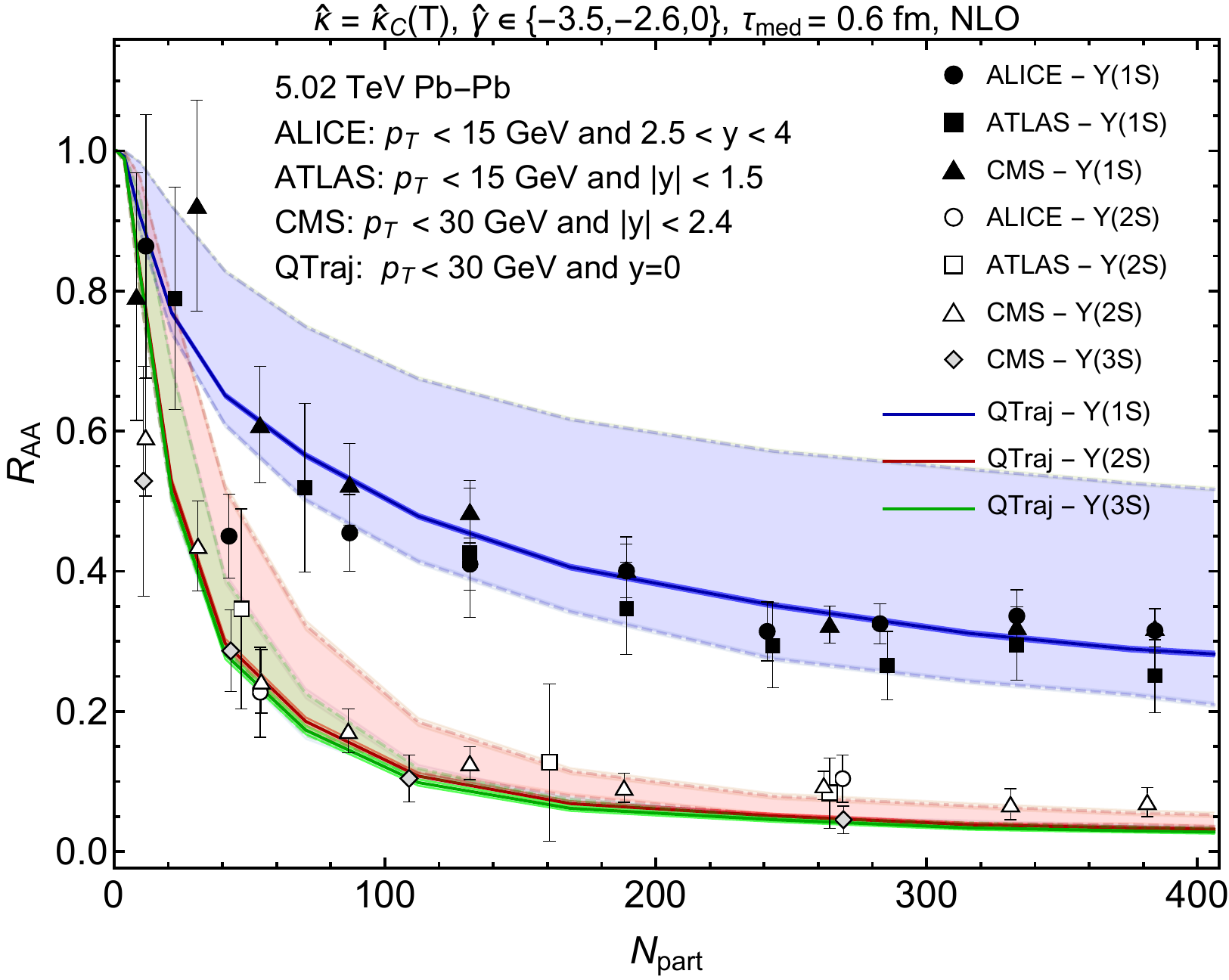}  \hspace{1mm}
\end{center}
\vspace{-5mm}
\caption{
The nuclear suppression factor, $R_{AA}$, for $\Upsilon(1S,2S,3S)$ as a function of the number of participants, $N_{\rm part}$.  The left figure shows variation of $\hat\kappa \in \{ \kappa_L(T), \kappa_C(T), \kappa_U(T) \}$ and the right figure shows variation of $\hat\gamma$ in the range $-3.5 \leq \hat\gamma \leq 0$.  In both cases, the solid line corresponds to $\hat\kappa = \hat\kappa_C(T)$ and the best fit value of $\hat\gamma = -2.6$.  Experimental results are from the ALICE~\cite{Acharya:2020kls}, ATLAS~\cite{Lee:2021vlb}, and CMS~\cite{Sirunyan:2018nsz,CMS:2022rna} collaborations.  Figures are taken from Ref.~\cite{Brambilla:2022ynh}.
}
\label{fig:raavsnpart1}
\end{figure}

\begin{figure}[t]
\begin{center}
\includegraphics[width=0.48\linewidth]{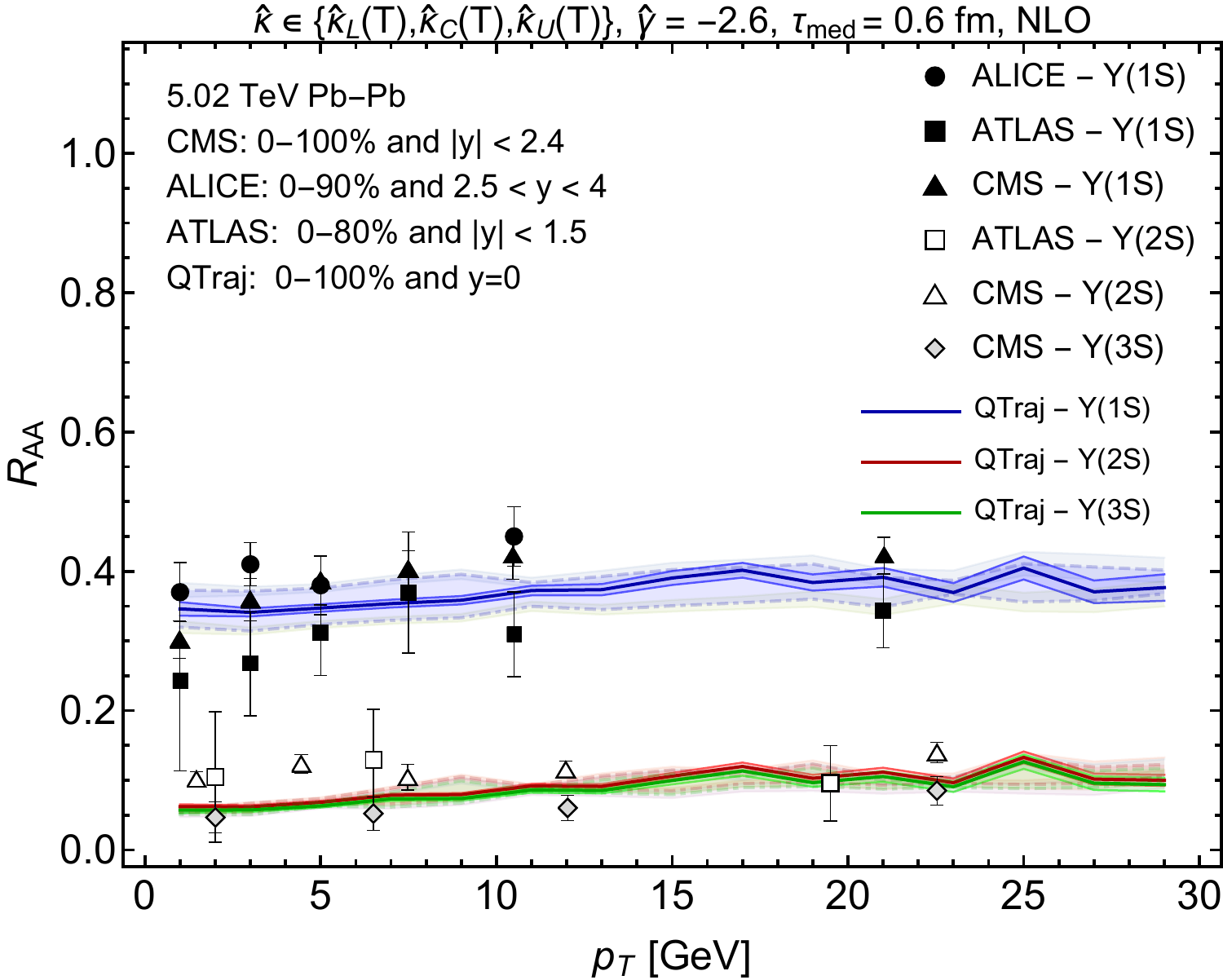} \hspace{2mm}
\includegraphics[width=0.48\linewidth]{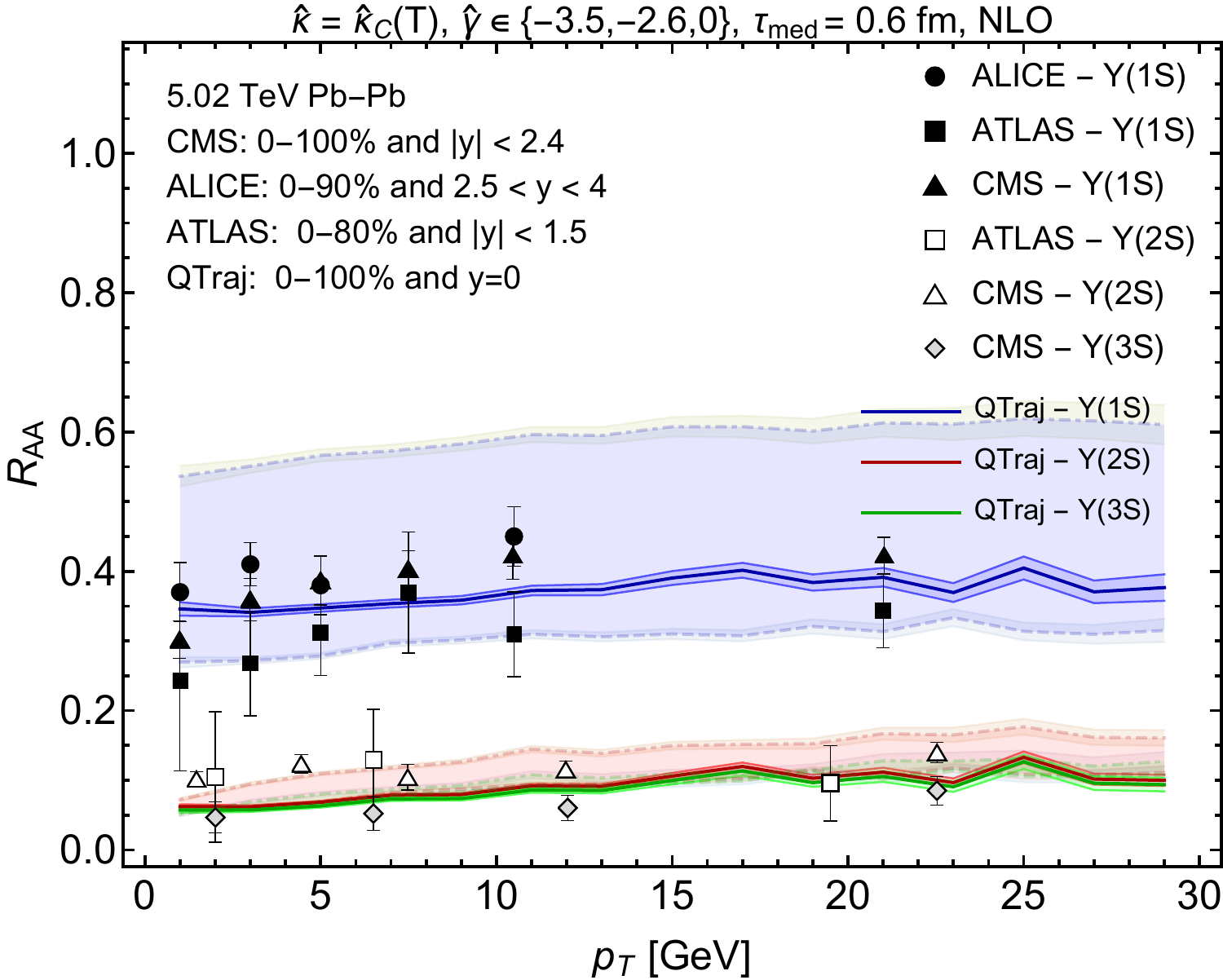} \hspace{1mm}
\end{center}
\vspace{-5mm}
\caption{
The nuclear suppression factor, $R_{AA}$, for $\Upsilon(1S,2S,3S)$ as a function of the transverse momentum, $p_T$.  The bands and experimental data sources are the same as Fig.~\ref{fig:raavsnpart1}.  Figures are taken from Ref.~\cite{Brambilla:2022ynh}.
}
\label{fig:raavspt1}
\end{figure}

\begin{figure*}[t]
	\centering
	\includegraphics[width=0.48\linewidth]{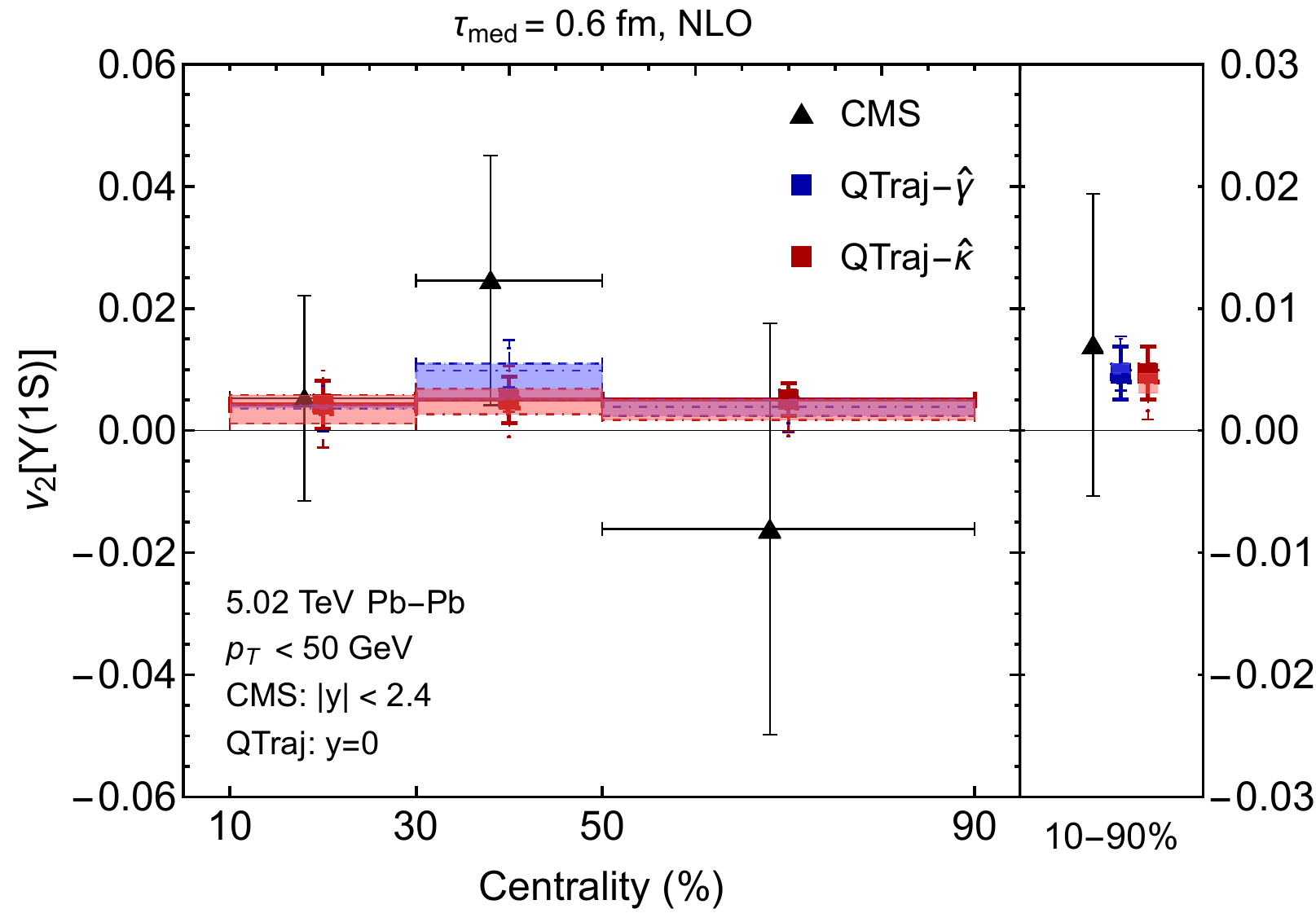} \hspace{2mm}
	\includegraphics[width=0.43\linewidth]{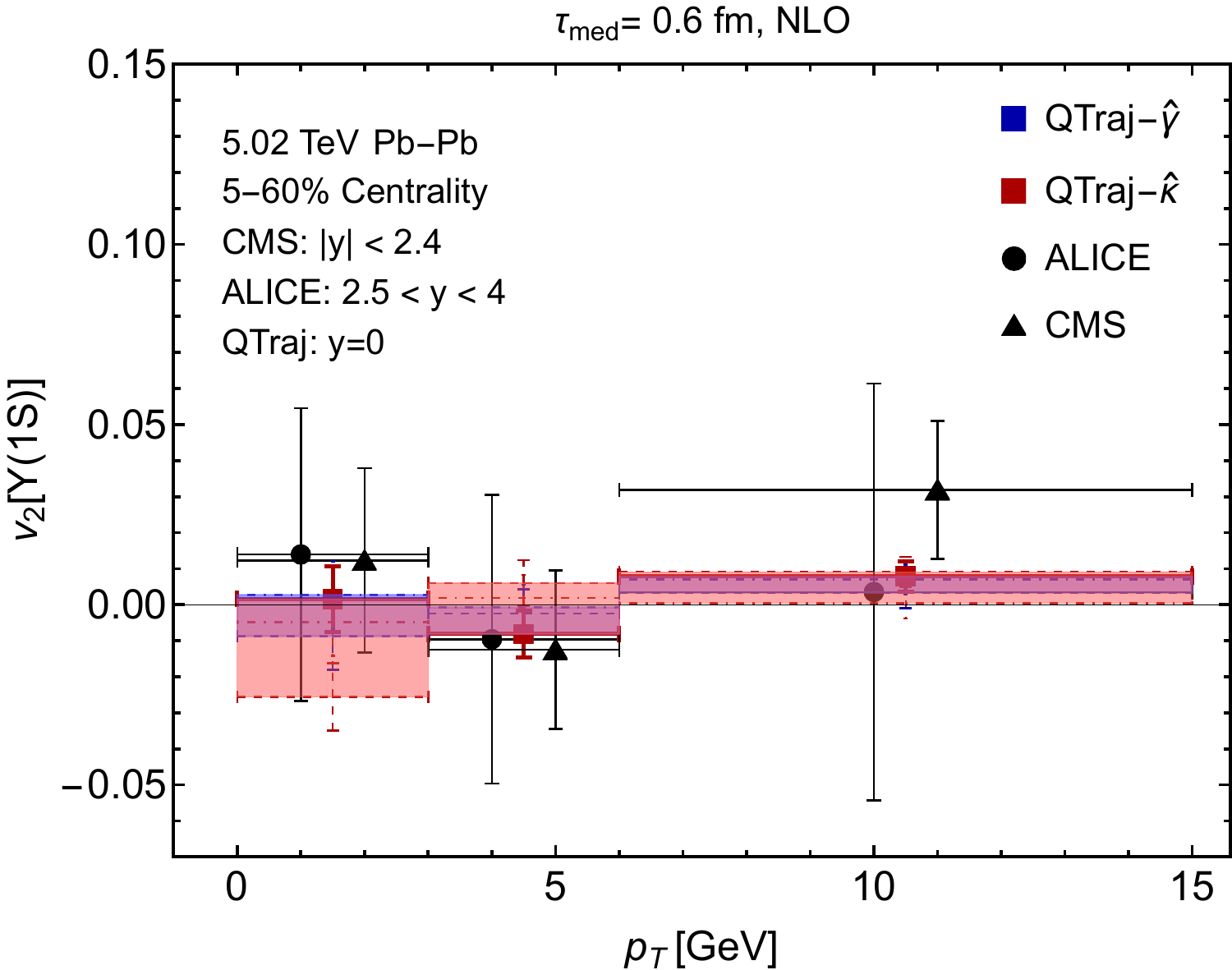} \hspace{1mm}
	\vspace{-2mm}
	\caption{The anisotropic flow coefficient $v_{2}$ as a function of centrality (left) and transverse momentum (right) obtained with IP-Glasma initial conditions.  We show the $\hat\gamma$ variation in blue and the $\hat\kappa$ variation in red and compare to experimental data from the ALICE and CMS collaborations \cite{ALICE:2019pox,CMS:2020efs}.  Figures are taken from Ref.~\cite{Alalawi:2022gul}.}
	\label{fig:v2}
\end{figure*}

In Figs.~\ref{fig:raavsnpart1} and \ref{fig:raavspt1}, we present our NLO predictions for $R_{AA}$ of the 1S, 2S, and 3S states as a function of $N_{\rm part}$ and $p_T$, respectively.    For these results we did not include the effect of dynamical quantum jumps.  In the left panel of Fig.~\ref{fig:raavsnpart1}, we show the variation of $\hat\kappa$ in the range $\hat\kappa \in \{ \hat\kappa_L(T), \hat\kappa_C(T), \hat\kappa_U(T) \}$ while holding $\hat\gamma = -2.6$.  
This value of $\hat\gamma$ was chosen as to best reproduce the $\Upsilon(1S)$ experimental data.  
In the right panel of Fig.~\ref{fig:raavsnpart1}, we show the variation of $\hat\gamma$ in the range $-3.5 \leq \hat\gamma \leq 0$ with $\hat\kappa(T) = \hat\kappa_C(T)$.  The solid line once again corresponds to $\hat\gamma = -2.6$.  
In both panels, the experimental data are from the ALICE~\cite{Acharya:2020kls}, ATLAS~\cite{Lee:2021vlb}, and CMS~\cite{Sirunyan:2018nsz,CMS:2022rna} collaborations.   We found that the NLO QTraj predictions without quantum jumps are in quite good agreement with the experimental data for $R_{AA}[1S]$ and $R_{AA}[3S]$.  However, for the 2S excited state, the NLO QTraj predictions without quantum jumps are somewhat lower than the experimental results, particularly for the most central collisions.  This over suppression could be due to having not included quantum jumps or could signal a breakdown of the pNRQCD framework for the excited states which are larger in radius.  We note, however, that as demonstrated in Fig.~\ref{fig:raavspt1}, when integrated over centrality, the $p_T$-dependence of $R_{AA}$ is quite reasonably reproduced.

In a recent paper \cite{Alalawi:2022gul} we also computed the NLO bottomonium elliptic flow $v_2[1S]$ using both smooth and fluctuating initial conditions for the hydrodynamic evolution.  In the case of fluctuating initial conditions, we used the MUSIC viscous hydrodynamics package \cite{Schenke:2010nt,Schenke:2011bn,MUSIC}, which includes the effects of both shear and bulk viscosity \cite{Ryu:2015vwa, Paquet:2015lta} and well reproduces a large set of soft-hadron observables, including identified hadron spectra and anisotropic flow coefficients.  For the fluctuating hydrodynamic initial conditions, we used fluctuating IP-Glasma initial conditions which incorporate the effects of the dense gluonic environment generated at early-times in a nucleus-nucleus collision \cite{Bartels:2002cj,Kowalski:2003hm,Schenke:2012wb}.  In Ref.~\cite{Alalawi:2022gul} we considered 2+1D boost-invariant evolution.  For further information about the hydrodynamical parameters, etc., we refer the reader to Ref.~\cite{Alalawi:2022gul}.  The hydrodynamic events used were sorted into centrality bins corresponding to 0-5\%, 5-10\%, 10-20\%, 20-30\%, 30-40\%, 40-50\%, 50-60\%, 60-70\%, 70-80\%, 80-90\%, and 90-100\%, with approximately 100 sampled hydrodynamical events in the 0-5\% and 5-10\% bins and 200 sampled hydrodynamical events in the other centrality bins.\footnote{The evolution files used are publicly available on Google Drive, \url{https://drive.google.com/drive/folders/1rEF7Cfe2DMHmZmlUyGvMW4RjM5BijqMt?usp=share_link}} As demonstrated in Ref.~\cite{Alalawi:2022gul}, for $R_{AA}$ the results obtained using fluctuating and smooth initial conditions are nearly identical, suggesting that initial state fluctuations do not have a large role to play in bottomonium suppression.

In Fig.~\ref{fig:v2}, I present the OQS+pNRQCD+IP-Glasma predictions for $v_{2}$ as a function of centrality (left panel) and transverse momentum (right panel) compared with experimental data from the ALICE and CMS collaborations \cite{ALICE:2019pox,CMS:2020efs}.  From Fig.~\ref{fig:v2} we see that the NLO OQS+pNRQCD+IP-Glasma framework predicts a rather flat dependence on centrality, with the maximum $v_2$ being on the order of 1\%.  In the right portion of the left panel, we present the results integrated over centrality in the 10-90\% range as two points that include the observed variations with $\hat\kappa$ and $\hat\gamma$, respectively.\footnote{The scale of the right portion of the left panel is different from the left portion of this panel.}  The size of the error bars reflects the statistical uncertainty associated with the double average over initial conditions and physical trajectories.  The red and blue shaded regions correspond to the uncertainty associated with the variation of $\hat\kappa$ and $\hat\gamma$, respectively.    Considering both variations, we find that, when integrated in the 10-90\% centrality interval and \mbox{$p_T < 50$ GeV}, the $v_2$ of the $\Upsilon(1S)$ is \mbox{$v_2[1S] = 0.005 \pm 0.002 \pm 0.001$}, with the first number corresponding to the statistical uncertainty and the second the systematic uncertainty associated with the variation of both $\hat\kappa$ and $\hat\gamma$.   Finally, in the right panel of Fig.~\ref{fig:v2}, I present the dependence of $v_2[1S]$ on transverse momentum.  We find that, at low momentum, there is a strong dependence on $\hat\kappa$, which could help to constrain this parameter.

\section{Conclusions}

In this proceedings contribution, I focused on recent works which apply the framework of OQS within the pNRQCD effective field theory.  I reviewed how to extend the OQS+pNRQCD framework and associated {\tt QTraj} code to NLO in the binding energy over the temperature.  This extension allows the framework to be applied at lower temperatures, making it a practical tool for studying bottomonium suppression in heavy-ion collisions.  I presented phenomenological predictions for the nuclear suppression factor, $R_{AA}$, and the elliptic flow coefficient, $v_2$, obtained using both smooth and fluctuating IP-Glasma initial conditions.  The impact of fluctuating initial conditions was found to be small when considering $R_{AA}$, however, a larger effect, albeit within statistical uncertainties, was observed for $v_2$.  For $R_{AA}[1S]$, $R_{AA}[3S]$, and $v_2[1S]$, we found quite good agreement between the NLO OQS+pNRQCD framework and experimental data given current theoretical and experimental uncertainties, however, we found that, irrespective of the hydrodynamic initial conditions used, the amount of $\Upsilon(2S)$ suppression is slightly overestimated.  Looking to the future, we plan to assess the full impact of dynamical quantum jumps and to compute higher anisotropic flow coefficients with fluctuating hydrodynamical backgrounds.  It would also be interesting to go beyond the underlying Coulomb potentials to assess the impact on excited states.

\vspace{2mm}

{\bf Acknowledgments}: I thank H. Alalawi, J. Boyd, N. Brambilla, M. Escobedo, A. Islam, C. Shen, A. Tiwari,  A. Vairo, P. Vander Griend, and J. Weber for their collaboration.  My work was supported by the U.S. Department of Energy, Office of Science, Office of Nuclear Physics Award No.~DE-SC0013470 and the Excellence Cluster ORIGINS at the Technical University Munich.

\bibliography{strickland}

\end{document}